\documentclass[aps,pra,twocolumn,superscriptaddress,showpacs]{revtex4-1}
\usepackage{amsmath}
\usepackage{bm}
\usepackage{graphicx}

\begin{document}

\title{Dicke and Fano effects in single photon transport}

\author{Maritza Ahumada, Natalia Cort\'es, M.\ L.\ Ladr\'on de Guevara, and P.\ A.\ Orellana}
\affiliation{Departamento de F\'{\i}sica, Universidad Cat\'{o}lica
del Norte, Casilla 1280, Antofagasta, Chile}

\begin{abstract}

The single photon transport through a one dimensional array of cavities
is studied theoretically. Analytical expressions of the reflection
and transmission are given. The transmission shows an energy
spectrum with forbidden and allowed bands that depends on the
detuning parameter of the system. We show that the allowed miniband
is formed due to the indirect coupling between the atoms in each cavity.
In addition, the band edges can be controlled by the degree of
detuning of the level of the atoms. We discuss the analogy between
this phenomenon with the Fano and Dicke effects.

\end{abstract}


\maketitle

\section{Introduction}

In the last years the single photon transport in one dimensional waveguide scattering by atoms or quantum dots has attracted great interest both theoretically and experimentally \cite{kim,Wallraff,Smolyaninov,Falk,Birnbaum,chang,jkim,Dayan}. The single photon transport has potential application in optical communication, optical quantum computer, quantum information, and quantum devices and as photon transistors. The possibility of implementing  photon transistors in analogy to the electron transistor  is a great challenge. Several kinds of waveguides have been involved, such as metal nanowires \cite{Ditlbacher}, silicon wires \cite{Zhu}, quantum resonators, etc. The behavior of photons in quantum resonator arrays is in analogy to the movement of electrons in periodic potentials. The photon can feel the presence of the quantum resonators and it can be transmitted through allowed mini bands and reflected by forbidden mini bands. The single photon scattering properties in the coupled resonator array waveguide embedded with a two-level or three-level atom have also been explored recently \cite{Gong,yue,Zhou}. In this context, recently Yue Chang {\em et al}.\cite{yue} considered new quantum devices based on a quantum resonator array. In these works, the authors have proposed a setup based on the coupled-resonator array with doped atoms, which is expected to exhibit perfect reflection within a wide spectrum of frequencies, and thus can perfectly reflect an optical pulse, or namely a single photon wave packet. They used a thick atomic mirror, which is made of an array of two-level atoms individually doped in some cavities arranged in a coordinate region of the one-dimensional coupled-cavity waveguide. 

Quantum interference effects as Fano effect play a crucial role in single photon transport \cite{fano}. Very recently Bei-Bei Li et al.\cite{beibei} experimentally observed Fano resonances in a single toroidal micro-resonator, in which two modes are excited simultaneously through a fiber taper. By adjusting the fiber-cavity coupling strength and the polarization of incident light, the Fano-like resonance line shape can be controlled. 

In this work we report further progress along the lines indicated
above. We study the single photon transport through a system of a 1D
array cavity. We obtain analytical expressions of
the reflection and transmission. We show that the transmission displays an energy
spectrum with forbidden and allowed bands that depends on the
detuning parameter of the system. We show that an allowed
miniband is formed by the indirect coupling between the levels of the atom in
each cavity. In addition, the band edges can be controlled by the
degree of detuning of the level of the atoms. We discuss the
analogy between this phenomenon with the Fano and Dicke effects.

\section{Model}

\begin{figure}[h!!]
\centerline{\includegraphics[width=9.8cm,clip]{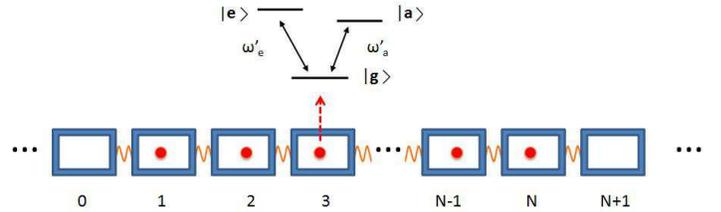}}
\caption{1D coupled cavity array with an embedded $N$ cavity-atom array
with one $V$-type three-level atom in each cavity.} \label{fig1}
\end{figure}
The system under consideration is shown schematically in Fig.~\ref{fig1}. This
consists in a 1D coupled cavity array with an embedded $N$ cavity-atom array
with one $V$-type three-level atom in each cavity. We denote by $|g\rangle$ the ground level and by $|a\rangle$ and $|e\rangle$
the two excited metastable levels. The atom is coupled to the cavity mode of frequency $\omega$ through the transitions
$g$-$a$ and $g$-$e$. The total Hamiltonian is described by the sum of a free atom
Hamiltonian of the atom  ${\mathcal{H}}_{a}$, the Hamiltonian describing the
propagation of a photon through the cavities ${\mathcal{H}}_{p}$, and the
term describing the atom-field interaction represented by a Jaynes-Cummings Hamiltonian under the rotating wave
approximation ${\mathcal{H}}_{I}$,
\begin{equation}
{\mathcal{H}}={\mathcal{H}}_{p}+{\mathcal{H}}_{a}+{\mathcal{H}}_{I},
\end{equation}
\begin{equation}
{\mathcal{H}}_{p}=\sum_{i=-\infty}^{\infty}\omega\hat{c}_{i}^{\dag}\hat{c}_{i}+\sum_{i=-\infty}^{\infty}v(\hat{c}_{i}^{\dag}
\hat{c}_{i+1}+\hat{c}_{i+1}^{\dag}\hat{c}_{i})\,
\end{equation}
\begin{equation}
{\mathcal{H}}_{a}=\sum_{j=1}^{N}(\,\omega'_{a}|a_{j} \rangle\langle
a_{j}|+\omega'_{e}|e_{j}\rangle\langle e_{j}|\,),
\end{equation}
\begin{eqnarray}
{\mathcal{H}}_{I}&=&\sum_{j=1}^{N}[
g_{a}(\hat{c}_{j}|a_{j}\rangle\,\langle
g_{j}|+\hat{c}_{j}^{\dag}|g_{j}\rangle\,\langle a_{j}|)\notag \\
&+&g_{e}(\hat{c}_{j}|e_{j}\rangle\,\langle
g_{j}|+\hat{c}_{j}^{\dag}|g_{j}\rangle\,\langle e_{j}|)]\,
\end{eqnarray}
\noindent where $\hat{c}_{j}^{\dag}$ and $\hat{c}_{j}$ are,
respectively, the creation and annihilation operators of one photon
in the $j$-th cavity, $v$ is the coupling between the cavities,
$\omega^{\prime}_{a,b}$ are the energy transitions and $g_{a}$
($g_{e}$) is the coupling constant between the field and transition
$|g\rangle\rightarrow|a\rangle$ ($|g\rangle\rightarrow|e\rangle$).
In what follows we assume $g_{a}=g_{e}\equiv \tilde{g}$.

The stationary state can be written as
\begin{equation}
|E\rangle=\sum_{i=-\infty}^{\infty}u_{i}|\,1_{i},g\,\rangle+\sum_{j=1}^{N}\left[d_{a,j}|\,0,a_{j}\,\rangle
+d_{e,j}|\,0,e_{j}\,\rangle\right],
\end{equation}
\noindent where $u_{i}$ represents the amplitude to find a photon in
the $i$-th cavity and the corresponding atom in the ground state,
and $d_{a,j}$ ($d_{e,j}$) the probability amplitude to find no
photon in the cavity array and the atom in the $j$-th cavity in the
excited state $a$ ($e$), while the rest of the atoms keep in the
ground state.

From the eigenvalues equation ${\mathcal{H}}|E\rangle=E|E\rangle$ it
is obtained the following linear difference equations
\begin{equation}
(E-\omega)\,u_{j}=v\,(u_{j+1}+u_{j-1}), \label{fuera}
\end{equation}
for $j\leq 0$  and  $j> N$, and
\begin{subequations}
\label{e:diferencias}
\begin{eqnarray}
(E-\omega)\,u_{j}&=&v\,(u_{j+1}+u_{j-1})+\tilde{g}\,(d_{e,j}+d_{a,j}), \label{eq-a}\\
(E-\omega'_{e})\,d_{e,j}&=&\tilde{g}\,u_{j}, \label{eq-b}\\
(E-\omega'_{a})\,d_{a,j}&=&\tilde{g}\,u_{j}, \label{eq-c}
\end{eqnarray}
\end{subequations}
for $j=1,\dots,N$. From Eqs. (\ref{eq-b}) and (\ref{eq-c}) we obtain
\begin{equation}
d_{e,j}=\frac{\tilde{g}\,u_{j}}{E-\omega'_{e}}, \quad
d_{a,j}=\frac{\tilde{g}\,u_{j}}{E-\omega'_{a}}, \label{dj-uj}
\end{equation}
which inserted in Eq. (\ref{eq-a}) make Eqs. (\ref{e:diferencias})
reduce to the single equation
\begin{equation}
(E-\omega-\tilde{\varepsilon})\,u_{j}= v\,(u_{j-1}+u_{j+1}),\quad
j=1,\ldots,N, \label{newdif1}
\end{equation}
where
\begin{equation}
\tilde{\varepsilon}=\tilde{g}^{2}\left[\frac{2E-\omega'_{a}-\omega'_{b}}{(E-\omega'_{a})(E-\omega'_{e})}\right]
\label{newdif2}
\end{equation}
\noindent is the renormalized energy. Thus, the problem reduces to that of a linear chain of  $N$ sites with
effective energies $\tilde{\varepsilon}$. In order to study the
solutions of Eq.~(\ref{newdif1}), we assume that the photon is
described by a plane wave incident from the far left with unity
amplitude and a reflection amplitude $r$ and at the far right by
a transmission amplitude $t$. That is,
\begin{align}
u_{j}^{(k)}  &  =e^{ikj}+re^{-ikj}\ , & j  &  <1\ ,\nonumber\\
u_{j}^{(k)}  &  =te^{ikj}\ , & j  &  >N\ . \label{solut2}
\end{align}
Inserting Eqs. (\ref{solut2}) in Eq. (\ref{fuera}) we obtain the following dispersion
relation for the incident photon
\begin{equation}
E=\omega+2 v cos(k).
\end{equation}
From Eqs. (\ref{fuera}), (\ref{newdif1}), and (\ref{solut2}) we obtain an inhomogeneous system of equations
for the probability amplitudes $u_j$ ($j=1,\dots, N$), $r$ and $t$, leading to the following
expression for $t$
\begin{equation}
t=\frac{2ie^{-ikN}}{\Delta}\sin{k}\ ,
\end{equation}
\smallskip\noindent with $\Delta$ given by
\begin{subequations}
\begin{equation}
\small{
\Delta=e^{-ik}\,\frac{\sin\left({N+1}\right){q}}{\sin q}+2\,\frac{\sin{Nq}}
{\sin q}+e^{ik}\,\frac{\sin\left({N-1}\right){q}}{\sin q}\ ,}
\end{equation}

if $|(E-\omega-\tilde{\varepsilon})/2v|\leq 1$ and
\begin{equation}
\small{
\Delta=e^{-ik}\frac{\sinh({N+1}){\kappa}}{\sinh\kappa}
+2\,\frac{\sinh{N\kappa}}{\sinh\kappa}+e^{ik}\,\frac{\sinh\left(  {N-1}\right)
{\kappa}}{\sinh\kappa}\ ,}
\end{equation}
\end{subequations}
if $|(E-\omega-\tilde{\varepsilon})/2v|\geq 1$. The reflection and transmission probabilities are $R=\left\vert
r\right\vert ^{2}$ and $T=\left\vert t\right\vert ^{2}$. Then if
$|(E-\omega-\tilde{\varepsilon})/2v|\leq1$ we have

\begin{subequations}
\small{
\begin{align}
R & =\frac{\sin^2 (N q)(\cos q+\cos k)^2}{\sin^2(N q)(\cos k \cos q
+1)^2+[\sin k\sin q \cos(Nq)]^2} ,\nonumber\\
T & =\frac{1}{\cos^2(Nq)+[\sin(Nq)(1+\cos q\cos k)/(\sin q \sin
k)]^2},\label{g2}
\end{align}
} where $q=\cos^{-1}[-(E-\omega-\tilde{\varepsilon})/2v]$. We note
that these probabilities oscillate as a function of both $N$ and
$q$. On the other hand, if
$|(E-\omega-\tilde{\varepsilon})/2v|\geq1$ we get
\small{
\begin{align}
R &  =\frac{\sinh^2 (N\! \kappa)(\cosh \!\kappa+\cos \!k)^2}{\sinh^2(N\! \kappa)(\cos\!k \cosh\!\kappa +1)^2+[\sin\!k\sinh \!\kappa  \cosh(N\!\kappa)]^2} ,\nonumber\\
T &  =\!\frac{1}{\cosh^2\!(N\! \kappa)+[\sinh(N \!\kappa)(1+\cosh\!\kappa \cos \!k)/(\sinh \!\kappa \sin\! k)]^2},\label{g3}
\end{align}
}
\end{subequations}
where $\kappa=\cosh^{-1}[-(E-\omega-\tilde{\varepsilon})/2v]$. Then
in this energy region $T\sim e^{-2N\kappa}$ tends to zero when $N$
is large and $R$ tends to unity.

\section{Results}

To avoid the profusion of free parameters, for the sake of clarity
we set $\Omega=E-\omega$, the energies of the atoms as,
$\omega'_{e}=\omega_{0}-\Delta\omega  $, and
$\omega'_{a}=\omega_{0}+\Delta\omega$ and we express all energies
in units of $\gamma$ with $\gamma=g^{2}/2v$, hereafter.

In what follows, we consider the degenerate case, $\Delta \omega=0$,
for different values of $N$. Simple  expressions for the
transmission and reflection can be readily obtained for $N=1$. For
$N=1$, the reflection reduces to a Breit-Wigner line shape of
semi-width $\gamma$,
\begin{equation}
R  =\frac{\gamma^{2}}{(\Omega-\omega_0)^{2}+\gamma^{2}},
\end{equation}
and the transmission takes the form of a symmetrical Fano line shape,
\begin{equation}
T=\frac{(\Omega-\omega_0)^{2}}{(\Omega-\omega_0)^{2}+\gamma^{2}}=\frac{(\epsilon+q)^{2}}{\epsilon^{2}+1},
\label{fano}
\end{equation}
with $q=0$ and $\epsilon=(\Omega-\omega_{0})/\gamma$. Notice that
$T$ vanishes just at $\Omega=\omega_0$. This result can be
interpreted as follows. The photon has two path when going from left to right, a direct and indirect one. In the latter the photon is absorbed and emiited by the atoms. The
destructive interference between these two paths gives rise to the
Fano effect in this case.

\begin{figure}[ht]
\centerline{\includegraphics[width=85mm,angle=0]{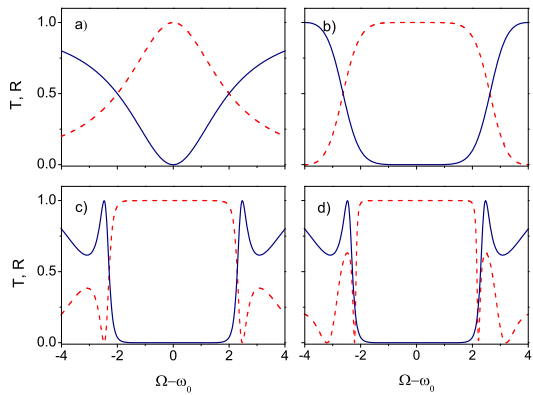}}
\caption{(Color online) Transmission (blue solid line) and reflection (red dash line) as a function of the detuning $\Omega-\omega_{0}$ for
$\Delta \omega= 0$ and a)~$N=1$, b)~$N=3$, c)~$N=5$ and d)~$N=7$.}
\label{fig2}
\end{figure}

Figure~\ref{fig2} shows the transmission and reflection
probabilities versus the detuning  $\Omega-\omega_{0}$ for different
values of $N$. As $N$ grows (Fig 2b-d) a forbidden mini band (gap)
is formed in $T$. It is apparent that $R$ ($T$)  tends to unity
(zero) within a range $[-2\gamma,2\gamma]$,  and the system behaves
as a quantum mirror within this interval of energies. 

\begin{figure}[ht]
\centerline{\includegraphics[width=85mm,angle=0]{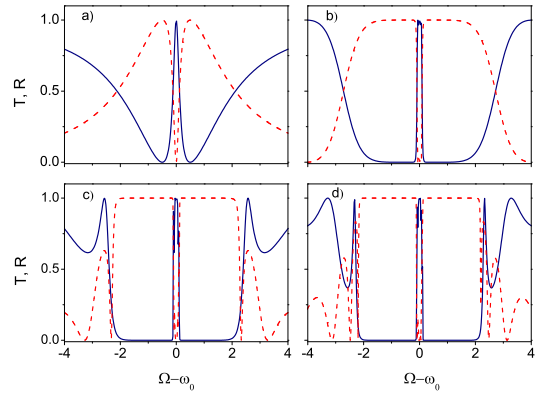}}
\caption{(Color online) Transmission (blue solid line) and reflection (red dash line) versus detuning for $\Delta = 0.5 $, for a) $N=1$, b) $N=3$,
c) $N=5$ and d)~$N=7$.} \label{fig3}
\end{figure}

Now, let us consider the situation with $\Delta \omega \neq 0$.
Figure~\ref{fig3} displays the reflection and transmission  vs the
detuning for different values of $N$. We note that an allowed band
develops at the center of the gap and the system becomes
transparent. Moreover the transmission  becomes always unity at the
center of the allowed band, independently of the value of $N$. This
behavior is analogue of the electromagnetically-induced transparency
– which has recently been studied theoretically and experimentally
for the structures based on micro- ring resonators and photonic
crystal cavities\cite{EIT,ring}.

\begin{figure}[h]
\centerline{\includegraphics[width=80mm,angle=0]{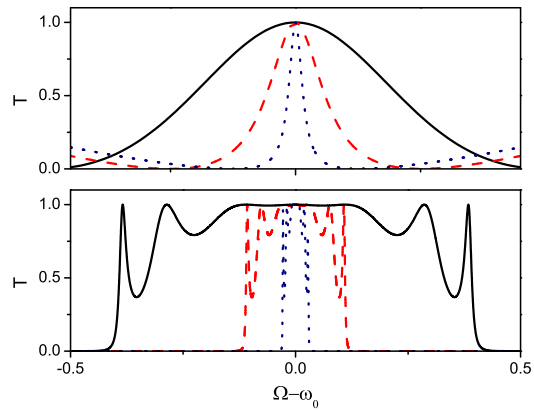}}
\caption{(Color online) Transmission  versus $\Omega$ for $\Delta = 1$ (black solid line)
$\Delta=0.5$ (red dash line) and $\Delta =0.25 $ (blue dotted line), for $N=1$ (upper panel) $N=7$ (lower panel).}
\label{fig4}
\end{figure}

Figure \ref{fig4} displays a zoom of the transmission at the center of the
band for $\Delta \omega =0.5\gamma$, for two values for $N$, $N=1$ (upper panel) and $N=7$ (lower panel).  We note the dramatic change of the width of the miniband as $\Delta \omega$ decreases. In fact, for $N=1$, it is
straightforward to show that for $\Delta \omega \ll\gamma$ the
width of this allowed transmission band is $\Delta \omega^{2}/2\gamma$,  the reflection can be written as a difference between two Breit-Wigner line-shapes with with $\gamma$ and $\delta$, respectively, and the transmission
can be written as a superposition of a symmetrical Fano line-shape and a Breit-Wigner line-shape,

\begin{eqnarray}
R  &\approx& \frac{\gamma^{2}}{(\Omega-\omega_0)^{2}+\gamma^{2}}-\frac{\delta^{2}}{(\Omega-\omega_0)^{2}+\delta^{2}},\\
T  &\approx& \frac{(\Omega-\omega_0)^{2}}{(\Omega-\omega_0)^{2}+\gamma^{2}}+\frac{\delta^{2}}{(\Omega-\omega_0)^{2}+\delta^{2}}.
\label{dicke}
\end{eqnarray}
\noindent
with $\delta=\Delta \omega^{2}/2\gamma$.

The above phenomenon resembles the Dicke effect, which takes place
in the spontaneous emission of a pair of atoms radiating a photon
with a wave length much larger than the separation between them
\cite{dicke,Pan}. The luminescence spectrum is characterized by a
narrow and a broad peak, associated with long and short-lived
states, respectively. The former state, coupled weakly to the
electromagnetic field,
is called \emph{subradiant}, and the latter, strongly coupled, \emph{%
superradiant} state. In the present case this effect is due to
indirect coupling between the atom states through the common cavity.
The states strongly coupled to the continuum give a forbidden
miniband with width $4\gamma$ and the states weakly coupled to the
continuum give an allowed Dicke miniband with width $\delta$. This
effect is a special case the Fano-Feshbach resonances in the systems
exhibiting more than one resonance\cite{feshbach}.

A physical realization of our set up may be made in a metal nanowire coupled to  quantum dots. In this realization the metal nanowire plays the role of 1D continuum and the region where the quantum dots are coupled to the nanowire.

\section{Summary}

In this work we have studied the transport of a single photon
through a system of a 1D array of cavities. We have obtained analytical
expressions of the transmission and reflection. We have shown that
the transmission displays an energy spectrum with forbidden and
allowed bands that depend on the detuning parameter of the system.
We have shown that the allowed miniband is formed by the indirect
coupling between the levels of the atom in each cavity. In addition,
the band edges can be controlled by the degree of detuning of the
level of the atoms. We have discussed the analogy between this
phenomenon with the Fano and Dicke effects. This setup seems a
suitable system to study these effects in experiments on single
photon transport.

\begin{acknowledgments}

The authors would like to thank financial support from FONDECYT  under Grants No. 1080660 and No. 1100560.

\end{acknowledgments}

\end{document}